\documentclass[12pt]{article}
\usepackage{amsmath,amsopn,amssymb,amsthm,amsfonts}
\usepackage[T2A]{fontenc}
\usepackage[cp1251]{inputenc}
\usepackage{graphicx}
\numberwithin{equation}{section}

\newtheorem{theorem}{Theorem}[section]

\newtheorem{definition}[theorem]{Definition}

\def\R {\mathbb{R}}

\def\min {{\rm min}}

\begin{document}

\vspace{7mm}
\centerline{\large \bf Wavelet analysis in problems of classification of ECG signals}
\vspace{3mm}
\centerline{\large \bf N.~K.~Smolentsev and P.~N.~Podkur}
\vspace{7mm}

\begin{abstract}
In this paper, the wavelet analysis is used to study the ECG signal.
We show that the high-frequency wavelet components of the ECG signal contain information on the functioning of the heart and can be used in diagnosis.
We describe the automated classification system that separates the ECG of sick and healthy persons using only a high-frequency ECG component.
\end{abstract}

\section{Introduction}
Heartbeat represents a complex electrochemical process. It is registered in the form of electrocardiogram by skin electrodes placed in certain places of body surface. One heartbeat cycle recorded on ECG usually consists of several bursts: P wave, then QRS complex, T wave and U wave (Fig. 2). After a while this complex of PQRSTU waves repeats. The specified waves, their sizes, sort, rhythms, intervals and PR and QT segments traditionally serve for heart diseases diagnosis. The form of waves, waves and segments' complexes duration, lengths variability of various cardiosignal intervals are analyzed. For ECG studying various statistical methods, Fourier transformation and spectrum analysis are used. More modern methods are based on wavelet analysis and artificial neural networks application \cite{Aldroubi}, \cite{Nagendra}, \cite{Pavlov}, \cite{Sambhu} and \cite{Shahbahrami}.
In these researches it is considered that high cutoff frequency of normal (without loading) cardiosignal noticeably influencing its form does not exceed 100 Hz. That is why in such analysis ECG of higher than 100 Hz frequency is almost not considered. Moreover, for cardiogram smoothing high-frequency components are usually deleted by means of various filters. It is clear, that at that part of information registered by the cardiograph is lost.
The physical origin of high frequencies of a cardiosignal is not clarified up to the end. They can include both the hardware noise and high-frequency physiological rhythms which are to a large extent consequence of heart electrical activity, as they are registered by the sensors located near heart. In the modern technical means the hardware noise are almost insignificant in comparison with physiological rhythms. That is why ECG high-frequency components reflect electrical heart activity and therefore for registration of high frequencies electrocardiographs of high resolution with sampling frequency of 5, 10, and 20 kHz are used nowadays.
Effective allocation of high-frequency components is possible with the use of wavelet decomposition of signal.
There are publications in which high frequencies of cardiosignal are analyzed by means of the continuous wavelet decomposition. For example, in works \cite{Ishikawa-02}, \cite{Ishikawa} and \cite{Mochimaru} the digitized cardiosignals with sampling rate of 5, 10 and 20 kHz and properties of cardiosignal at frequencies from 25 Hz to 400 Hz were studied. In work \cite{Smolen} frequency and stochastic characteristics of high-frequency cardiosignal components with the use of discrete wavelet decomposition have been analyzed.

Long ECG records of patients contain a huge amount of data. That is why detection of disease symptoms is time-consuming process which demands the detailed analysis of all ECG data length. Reliable automatic classification and system detecting ECG parameters' anomalies would guarantee more exact diagnostics and significant facilitation of cardiogram "decoding". Therefore creation of the automated system detecting ECG parameters' anomalies and ECG classification is an urgent task. In this direction certain results based on studying of low-frequency ECG characteristics are achieved, e.g. \cite{Nagendra}, \cite{Sambhu} and \cite{Shahbahrami}. Nowadays, the problem of creation of the automated systems of ECG classification considering high-frequency components is open issue.

In this paper a new design method of the automated classifying system for electrocardiograms recognition (ECG) of healthy and sick persons, based only on high-frequency components of ECG signal with the use of statistical images recognition is offered. Cardiograms of two groups of patients were studied: healthy and those who came through myocardial infarction. The first step of classification method is ECG wavelet decomposition to the 4th level and allocation of four high-frequency ECG components.
The choice of the 4th decomposing level is explained by the fact that the first four high-frequency components represent high ECG frequencies from 30 to 350 Hz, and low-frequency component represents the undistorted smoothed ECG signal cleared of high-frequency oscillations. In case of more deep signal expansion the following high-frequency component has frequency spectrum to 30 Hz, and low frequency component is significantly distorted.
For each of the first four components of wavelet decomposition there is a number of ECG numerical signs, including energy, entropy and frequency characteristics, 21 signs in total. During the second step reduction of the dimension of the feature space by using scatter matrix is made for two chosen ECG groups. It has turned out that the reduced feature space is one-dimensional.
Histograms of values of this one-dimensional feature for groups of healthy and sick persons are constructed.
The third step is finding of the dividing constant which is able to distinguish both groups of ECG records. For testing 96 ECG records of patients with normal cardiograms and 120 ECG records of the patients who came through myocardial infarction are used. Only three features (3\%) of 96 given features values of the first group are referred by the classifier to patients group and only 20 features (< 17\%) of 120 given features values of the second patients group are referred by the classifier to ECG group of healthy persons. Considering that for each patient the system determines 12 features by 12 standard assignments, testing results show well classification accuracy.

\section{Materials and methods}
\subsection{Materials}
For classification system creation and its testing ECG data sets of two groups are studied: healthy persons with normal ECG data and patients who recently came through myocardial infarction (MI). For the analysis digitized 30 seconds long cardiosignals made on the high-resolution cardiograph (1028 counts per second) "Cardiotekhnika – 4000, by EcgShell" were used. The cardiosignal is registered on 8 standard channels:
$L$ -- left hand $(+)$ and right hand $(-)$, $F$ -- left leg $(+)$ and right hand $(-)$ and six chest leads marked as $C_1 - C_6$. Of 8 cardiograph channels $L$, $F$, $C_1, C_2,\dots , C_6$ there are 12 so called standard leads \cite{Vorobyov}: $I$, $II$, $III$, $aVR$, $aVL$, $aVF$, $V_1, V_2,\dots , V_6$ according to formulae:
$$
I = L,\quad II = F,\quad III = F -L, \quad aVR = -(L+F)/2,\quad aVL = L -F/2, \quad  aVF =F -L/2,
$$
$$
V_i = C_i -(L+F)/3,	\quad i = 1,2,\dots , 6.
$$
8 seconds long fragments were chosen from ECG records of each of 12 leads, wavelet decomposition and features calculation are made for them. All calculations are performed in MATLAB system \cite{Smolen-ML} using wavelet analysis package MATLAB Wavelet Toolbox. Functions of this wavelet analysis package provide correct processing of boundary values at filters action by symmetric signal continuation. Thus, for each patient 24 ECG fragments were studied. During the construction of classifying system ECG records of two groups of patients were used. The first group of ECG records of healthy persons contains 96 ECG fragments for four persons aged from 21 to 27. The second group of ECG records of the patients who recently came through myocardial infarction (subacute period) contains 120 ECG fragments for five patients aged from 44 to 55. For testing 96 ECG  fragments of four healthy persons aged from 21 to 56 years and 120 ECG fragments for five patients aged from 45 to 57 years which recently came through myocardial infarction (subacute period) were used. All data contain only ECG records and information about patient's age. Cardiograms' analysis for the purpose of the diagnosis was not made.

\subsection{Methods}

\subsubsection{Wavelet-decomposition}
The key elements of wavelet-analysis are two functions: the scaling function  $\varphi(t)$ and the wavelet function $\psi(t)$, satisfying equations
\begin{equation}\label{varphi}
\varphi(t)=\sqrt{2}\sum_{k\in \mathbb{Z}} h_k \varphi(2t-k),
\end{equation}
\begin{equation}\label{psi}
\psi(t)=\sqrt{2}\sum_{k\in \mathbb{Z}} g_k \varphi(2t-k),
\end{equation}
where, $h_k$ and $g_k$ are the low and high-pass filter coefficients, respectively.
Moreover, it is usually assumed that the set $\{\varphi(t-k); k\in \mathbb{Z}\}$ is an orthonormal basis of a subspace $V_0\subset L^2(\R)$.

Let $V_j$ be the closed subspace in $L^2(\R)$ generated by the functions $\{\sqrt{2^j}\varphi(2^j t-k); k\in \mathbb{Z}\}$, $j\in \mathbb{Z}$.
Then from equation (\ref{varphi}) follows
\begin{equation}\label{MultiR}
\dots \subset V_{-1} \subset V_{0}\subset V_{1}\subset \dots \subset V_{j}\subset V_{j+1}\subset \dots
\end{equation}
Thus, the scaling function $\varphi$ generates a multiresolution representation in $L^2(\R)$.

\begin{definition}
(Multiresolution Representation).We define a multiresolution representation in $L^2(\R)$ as a sequence of closed subspaces $V_j$, $j\in \mathbb{Z}$, of $L^2(\R)$, satisfying the following properties:
\begin{enumerate}
  \item $V_j\subset V_{j+1}$.
  \item $f(t)\in V_j$ if, and only if, $f(2t)\in V_{j+1}$.
  \item $\cup_{j\in \mathbb{Z}}V_j =\{0\}$.
  \item $\overline{\cap_{j\in \mathbb{Z}}V_j} =L^2(\R)$.
  \item The set $\{\varphi(t-k); k\in \mathbb{Z}\}$ is an orthonormal basis of $V_0$.
\end{enumerate}
\end{definition}
The functions
$$
\varphi_{j,k}(t)= \sqrt{2^j}\varphi(2^j t-k),\quad k\in \mathbb{Z}
$$
form an orthonormal basis of the subspace $V_j$.
Then the representation of a signal $f(t)$ in the scale $V_j$ is given by the formula
$$
A_j(f) =\sum_{k=-\infty}^{\infty}\langle f,\varphi_{j,k}\rangle \varphi_{j,k}(t).
$$
The coefficients
$$
a_{j,k} = \langle f,\varphi_{j,k}\rangle, \quad k\in \mathbb{Z}
$$
are called the \emph{approximation coefficients} (or low-pass coefficients) of the function $f(t)$ on the scale $V_j$.

In the limit, $j\rightarrow\infty$, the scaling function tends to the Dirac delta function; hence the corresponding low-pass coefficient $a_{j,k}$ tends to the value of the function at location $k$.
This allows us to take its representation $A_j(f)$ in the scale $V_j$ instead of the function $f(t)$ for a sufficiently large $j = j_0$. Instead of the values of the function $f(t)$, we can consider its approximation coefficients $a_{j,k}$ (for a sufficiently large $j = j_0$).

For every $j\in \mathbb{Z}$ we have $V_{j-1}\subset V_j$.
Let $W_{j-1}$ be an orthogonal complement to $V_{j-1}$ in the space $V_{j}$. We have
\begin{equation}\label{V-W}
V_j=V_{j-1}\oplus W_{j-1}.
\end{equation}
The wavelet functions
$$
\psi_{j-1,k}(t)= \sqrt{2^j}\psi(2^{j-1} t-k),\quad k\in \mathbb{Z}
$$
form an orthonormal basis of the subspace $W_{j-1}$.
Then the representation of a signal $f(t)$ in the scale $V_j$ is given by the formula
$$
A_j(f) =\sum_{k=-\infty}^{\infty}\langle f,\varphi_{j-1,k}\rangle \varphi_{j-1,k}(t) + \sum_{k=-\infty}^{\infty}\langle f,\psi_{j-1,k}\rangle \psi_{j-1,k}(t).
$$
The coefficients
$$
d_{j-1,k} = \langle f,\psi_{j-1,k}\rangle, \quad k\in \mathbb{Z}
$$
are called the \emph{coefficients of details} (or high-pass coefficients) of the function $f(t)$ in the scale $W_{j-1}$.

Thus we obtain two representations:
$$
A_j(f) =\sum_{k=-\infty}^{\infty}a_{j,k} \varphi_{j,k}(t) = \sum_{k=-\infty}^{\infty}a_{j-1,k} \varphi_{j-1,k}(t) + \sum_{k=-\infty}^{\infty}d_{j-1,k} \psi_{j-1,k}(t).
$$
The coefficients $a_{j-1,k}$ and $d_{j-1,k}$ are expressed in terms of $a_{j,k}$ by formulas
\begin{equation}\label{Wav-Dec}
a_{j-1,k}=\sum_n h_n a_{j,n+2k}, \qquad
d_{j-1,k}=\sum_n g_n a_{j,n+2k}.
\end{equation}

You can repeat the wavelet decomposition procedure (\ref{V-W}) $N$ times.
Then we obtain the decomposition of $V_j$ into an orthogonal direct sum
\begin{equation}\label{V-W-W}
V_j=V_{j-N}\oplus W_{j-N}\oplus W_{j-N+1}\oplus\dots \oplus W_{j-1}.
\end{equation}
If initially the signal was represented by the coefficients $a_{j,k}$, now we have obtained the approximation coefficients $a_{j-N,k}$ on a smaller scale $V_{j-N}$ and the set of high-frequency coefficients $d_{j-s,k}$:
$$
f(t) \mapsto \{a_{j,k}\} \mapsto \{\{a_{j-N,k}\}, \{d_{j-N,k}\}, \{d_{j-N+1,k}\},\dots , \{d_{j-1,k}\} \}.
$$
This is the discrete wavelet-transform. For more details, see \cite{Daube} and \cite{Smolen}.

If the initial signal $f(t)$ is discrete, then its values $S_k=f(\Delta \cdot k)$, $k\in \mathbb{Z}$ can be taken as initial coefficients, $a_{j,k}=S_k$.
In this case, the array of values $S = \{S_k\}$ is decomposed into two arrays of coefficients:$D_1 = \{d_{1,k}\}$ and $A_1 = \{a_{1,k}\}$,
\begin{equation} \label{dec}
a_{1,k}=\sum_n h_n S_{2k+n}, \qquad d_{1,k}=\sum_n g_n S_{2k+n}.
\end{equation}
The array $A_1$ represents smoothed part of signal and it is called \emph{approximation coefficient} array. The $D_1$ array represents \emph{details} in which the initial signal $S$ differs from its smoothed part.
The result of filter $\{h_n\}$ action is low-frequency approximation of the signal.
The result of filter action $\{g_n\}$ is high-frequency part of the signal.
The filters $\{h_n\}$ and $\{g_n\}$ are also used for signal reconstruction $S =\{S_n\}$ using the formula:
\begin{equation} \label{rec}
S_n=\sum_n \left(h_{n-2k} a_{1,k} +g_{n-2k} d_{1,k}\right).
\end{equation}

During multi-level wavelet analysis the procedure of wavelet decomposition (\ref{rec}) is used many times to approximation coefficient array. It can be represented schematically as follows (fig. 1):
\begin{figure}
  \centering
  \includegraphics[width=120mm]{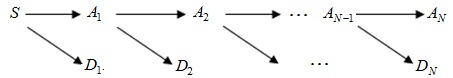}\\
  \caption{Multi-level wavelet decomposition of signal $S$}\label{1}
  \label{wav-dec}
\end{figure}

Reconstruction of the initial signal is made consistently in reverse order.
If we apply the reconstruction procedure to only one set of coefficients, and all other coefficients are taken as zero, then we obtain a part of the signal corresponding to one set of coefficients.
We will call this part the \emph{component of a signal}.
The components of the signal, reconstructed only by coefficients of details $D_1$, $D_2$, ..., $D_N$, will be called \emph{high-frequency components} of signal $S$ and will be denoted as $ RecD_1 $, $ RecD_2 $, ..., $ RecD_N $, respectively (Fig. 2).

\begin{figure}
  \centering
  \includegraphics[width=160mm]{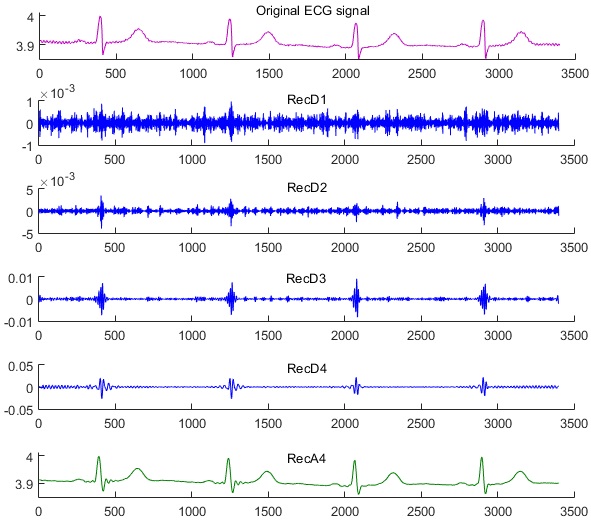}\\
  \caption{Wavelet components of ECG signal during decomposition to the 4th level (across is signal counting from 1 to 3400)}
  \label{wav-dec}
\end{figure}

For example, $RecD_2$ is the signal component, reconstructed on the following set of wavelet coefficients $\{0, D_2, 0, \dots, 0\}$, where 0 means the array from zeros. Similarly, low-frequency components $RecA_1$, $RecA_2$,\dots, $RecA_N$  came out by reconstruction only of one set of approximating coefficients. The sum of all signal components $RecD_1$, $RecD_2$, \dots , $RecD_N$  and $RecA_N$  is equal to the original signal:
$$
S = RecD_1 + RecD_2 + \dots + RecD_N + RecA_N.
$$

\subsubsection{Feature space}
For every high-frequency decomposition components $RecD_1$, $RecD_2$, \dots , $RecD_N$ many various statistical, frequency and stochastic characteristics can be calculated:
\begin{itemize}
  \item maximum absolute value;
  \item dispersion;
  \item $L_1$- and $L_2$-energy;
  \item relative $L_2$-energy;
  \item maximum value of the power spectrum;
  \item frequency, where maximum value of the power spectrum is achieved;
  \item Shannon's entropy;
  \item the Hurst exponent and other characteristics of randomness;
  \item average value of instantaneous frequency of oscillations calculated on the basis of discrete Hilbert transform
\end{itemize}
Definitions of the listed parameters will be reminded. $L_1$-energy of signal $X = \{x_n\}$ is the sum of elements’ modules of $x_n$, and $L_2$-energy of signal $X$ is the sum of squares of elements’ modules of $x_n$. Relative energy of signal component is the ratio of $L_2$-energy of component to $L_2$-energy of the entire signal.
Discrete Fourier transform $C = {\rm fft}(X)$ of signal $X = \{x_n\}$ of length $N$ is made according to the formula:
$$
c_k=\sum_{n=0}^{N-1}x_n e^{-i\,\frac{2\pi}{N}kn}, \qquad k=0,1,\dots, N-1.
$$
The obtained signal $C =\{c_k\}$ shows frequency properties of the array $\{x_n\}$, therefore it is called a \emph{signal spectrum} $\{x_n\}$. As values $\{c_k\}$  can be complex, the so-called \emph{power spectrum} of frequencies (frequency spectrum) calculated according to the formula is of interest:
$$
P_k=\frac{|c_k|^2}{N}, \qquad k=0,1,\dots, N-1.
$$
The diagram of power spectrum $P_k$ (Fig. 3) is usually figured for values $k$  in the range from $0$ to the middle of $N/2$ as it is symmetric and it makes sense to consider only the frequencies which are smaller than Nyquist-frequencies corresponding to $k =N/2$. In figure 3 diagrams of power spectra for components of cardiosignals of two patients from different groups are given.
As power of frequencies over 360 Hz are almost not noticeable in the figure, these diagrams are given in the range from 0 to 360 Hz.

\begin{figure}
  \centering
  \includegraphics[width=150mm]{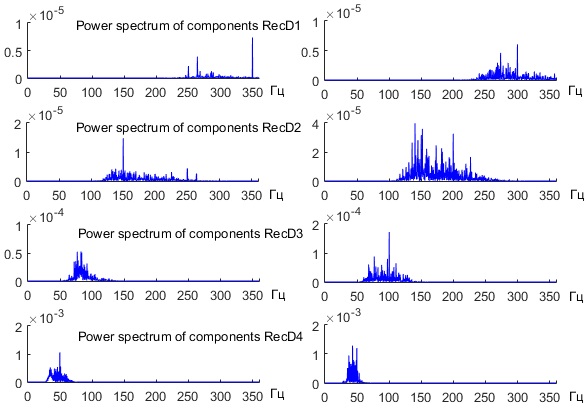}\\
  \caption{Power spectrum diagrams of wavelet components of a ECG signal when decomposition is to the 4th level on the interval from 0 to 360 Hz. On the left is for healthy patient, on the right -- for sick patient}
  \label{wav-dec}
\end{figure}

The Shannon's entropy is defined by the formula $E(X)=-\sum_k\, x_k^2\, \ln(x_k^2)$.

The behavior of signal components can be rather difficult and of randomness nature.
The degree of randomness can be estimated by Hurst exponent. It represents a propensity score of process to trends. $H > 0.5$ value means that process dynamics directed to a certain side in the past most likely will lead to movement continuation in the same direction. If $H < 0.5$, then it is predicted that process will change its direction, and $H = 0.5$ means uncertainty. Calculation of Hurst exponent for the signal $\{x_n\}$ is usually made on the basis of the so-called $RS$ analysis. Let’s remember a classical $RS$ method for Hurst exponent finding. Let the signal $\{x_n\}$ of length $N$ be given. Then Hurst exponent H can be found from the ratio:
$$
R/S = (N/2)^H,
$$
where $S$ is a standard deviation with selective average $m_X$, and $R$ is the so-called range, accumulated deviation from average:
$$
R=\max_{1\le n \le N}\sum_{k=1}^n\, (x_k-m_X) -\min_{1\le n \le N}\sum_{k=1}^n\, (x_k-m_X).
$$
In MATLAB there is \texttt{wfbmesti} function which assesses the fractal Hurst index of H signals. Hurst exponent is one of characteristics of randomness degree. Besides, according to the available one-dimensional data it is possible to construct dynamic system in multivariate phase space for which the observed variable will be one of coordinates, and system tracks lies on some set having fractal structure and fractional dimension. Therefore for the assessment of randomness of wavelet-coefficients and ECG signal component it is also possible to use such characteristics as: phase space dimension, fractal dimension and correlation dimension. It is possible to read about it in more detail in the work \cite{Smolen} where these characteristics were used for the analysis of high-frequency components of ECG signal, see also \cite{Adeli} where stochastic parameters were used for research EEG of healthy persons in comparison with EEG of patients with epilepsy.

We will remind that Hilbert transform  $y(t) = H(x(t))$ of the function $x(t)$ is defined by the formula:
$$
y(t) = \frac{1}{\pi} \int_{-\infty}^{\infty}\frac{x(\tau)}{t-\tau}\, d\tau,
$$
of course, if this integral exists in sense of a principal value.
One of the basic properties of Hilbert transform is that $H(H(x)) = -x$.
Then complex function $z(t) = x(t) + i y(t)$ is eigen-function for Hilbert transform: $H(z(t)) = -i z(t)$. Let us write the function $z(t)$ in complex form $z(t) = A(t)\,e^{i\theta(t)}$. Amplitude $A(t)$ is defined as function module $z(t)$, and instantaneous frequency $\omega$ of oscillations -- by the formula $\omega = \frac{d\theta}{dt}$, where $\theta(t) = \arctan(y/x)$. In MATLAB there is a function $\texttt{z = hilbert(x)}$ for Hilbert transform of discrete signal $X$. Then formula
\vspace{2mm}
\centerline { \texttt{instfreq = Fs/(2*pi)*diff(unwrap(angle(z)));}}
\vspace{2mm}
gives instantaneous frequency of the signal.

For ECG studying it is possible to use all listed numerical signs or to choose some of them. Selected numerical characteristics of ECG decomposing components form a vector of features $Y = [y_1, y_2, \dots, y_n]$ in $n$-dimensional feature space.

\subsubsection{Reduction of the feature space}\label{Reduction}
Let us suppose that wavelet decomposition is done and feature vectors $Y = [y_1, y_2, \dots, y_n]$ are made for some set of ECG records. Features space can have too big $n$ dimension that complicates creation of classifiers, as it assumes working with high order matrixes.
It is desirable to somehow reduce its dimension without essential loss of information. It is accepted to carry out decrease of dimension by means of linear display of all space of features on some smaller subspace \cite{Fukunaga}, ch. 10. It is performed by $A$ matrix, in which the number of columns $m$ is less than number of lines $n$ so that the initial vector $Y$ after linear  decomposition $Z = A^T\,Y$ is projected onto the vector $Z$, which dimension $m$ is significantly less (for example, 2 or 3, then it is possible to visualize classifiers in two- or three-dimensional space).

The reduction matrix $A$ can be defined in several various ways. The main idea of these methods is in defining the direction in which dispersion of features' vector $Y$ is the biggest by means of covariance matrix analysis. This direction is considered to be the most informative \cite{Fukunaga}, ch. 10.

If $Y$ data consist of several classes (for example, ECG features of sick and healthy persons), then it is necessary to choose such subspace of the most informative features which is the most effective from the point of view of classes' divisibility. Discriminant analysis procedure which at the same time gives dimensions and divides classes is based on scatter matrixes \cite{Fukunaga}, ch. 10.

It can be described as follows: Let us suppose that we have $N$ vectors of $n$-dimensional features $Y_i$, $i = 1, \dots, N$, as $Y_i = [Y_{i1}, Y_{i2},\dots, Y_{in}]$. It is also supposed that elements of this data set can be divided into some number of $c$ classes.
In this case $n$ is a number of extracted features after the wavelet analysis of ECG signal and $N$ is a number of decomposed fragments of ECG data. In other words, all set of basic data $\{Y_i, i = 1, \dots, N\}$ can be divided into $c$ subsets $\{Y_i^{(k)},  i = 1,\dots, N_k, k = 1, \dots , c \}$.
In our case, the number of classes $c$ is equal to two: ECG of normal persons and sick patients.

Scatter matrix $S_w$ within the classes shows dispersion of features concerning vectors of expected value of classes
\begin{equation} \label{Sw}
S_w=\sum_{k=1}^c P_k E\left\{(Y^{(k)} -M_k)(Y^{(k)} -M_k)^T \right\} =\sum_{k=1}^c P_k \Sigma_k,
\end{equation}
where $P_k$ is probabilities of getting into a certain class within all data set, $E\{\,\}$ is the operator of the expected value, $M_k=E\{Y^{(k)}\}$ . In practice these values are approximated by selective estimates:
\begin{equation} \label{estimates}
M_k=\frac{1}{N_k}\sum_{j=1}^{N_k}Y_j^{(k)}, \,
\Sigma_k =\frac{1}{N_k}\sum_{j=1}^{N_k} (Y_j^{(k)} -M_k)(Y_j^{(k)} -M_k)^T, \,
P_k=\frac{N_k}{N}, \, k = 1, \dots , c.
\end{equation}

Scatter matrix $S_b$ between classes shows vectors dispersion of expected values around average value of mixture and is defined as follows:
\begin{equation} \label{Sb}
S_b=\sum_{k=1}^c P_k (M_k -M_0)(M_k -M_0)^T, \quad M_0 =\sum_{k=1}^c P_k M_k .
\end{equation}

To obtain criterion of classes divisibility and the choice of optimum features, some number which increases at increase of dispersion between classes or at reduction of dispersion within the class is linked to these matrixes. There are different approaches which consider these two requirements. In this article the following criterion is accepted \cite{Fukunaga}, ch. 10.:
\begin{equation} \label{J1}
J_1={\rm tr}\{{S_w}^{-1}S_b\} =\sum_{i=1}^m \lambda_i,
\end{equation}
where tr\{{\}} is a trace of square matrix and $\lambda_i$  are eigenvalues of the matrix ${S_w}^{-1}S_b$.
Now it is necessary to choose such subspace of features which maximizes criterion $J_1$. Let us consider the matrix $A = [\Psi_1\Psi_2\dots\Psi_m]$ created as a set of columns $\Psi_1,\Psi_2,\dots,\Psi_m$ which are eigenvectors of matrix ${S_w}^{-1}S_b$, which correspond to $m$ the biggest eigenvalues of this matrix: $\lambda_1\ge \lambda_2\ge\dots \ge \lambda_m\ge \lambda_{m+1}\ge \dots \lambda_n$.
Then transition from full space of features to the given space of features is made by projection $Z = A^TY$ on vectors $\Psi_1,\Psi_2,\dots,\Psi_m$ with the help of matrix $A$. Herewith some information is lost. Every eigenvector $\Psi_i$ bears the amount of information which corresponds to the value of the corresponding eigenvalue $\lambda_i$. Therefore the relative measure of the saved information can be calculated as follows: $\sum_{i=1}^m \lambda_i /\sum_{i=1}^n \lambda_i \cdot 100\%$.
Results of reduction of dimension with index of informational content which are higher than 85\% are considered as satisfactory.

Despite some loss of information, reduction of dimension improves divisibility between classes and facilitates the task of classification. Besides, reduction of dimension allows to visualize results. It is very difficult to analyze any classifier in 21-dimensional space of features (as in this work) and it is almost impossible to present it.

During reduction $Z = A^TY$ of features space the reducing matrix $A$ of dimension of $n$-on-$m$ mapping each of vectors $Y$ sets of basic data on the corresponding $m$-dimensional vector $Z$. Therefore, vectors $Z$ are also divided into $c$ subsets $\{Z_i^{(k)} = A^T Y_i^{(k)}$,  $i = 1, \dots , N_k$, $k = 1, \dots, c\}$. Now the task is in creation of classifiers, i.e. of functions which divide all these subsets.

\subsubsection{Linear classifiers}
The algorithm of classification assumes the division of patients into two groups (healthy and those who had myocardial infarction (MI)). Therefore it is possible to use linear classifiers. Let us remind their construction \cite{Fukunaga}, ch. 4, 10. The linear classifier has the form of linear heterogeneous function $h(Z) = V^TZ + v_0$, where $Z$ is $m$-dimensional vector of data received after dimension reduction, $V$ is vector of coefficients and $v_0$ is a constant term. If $h(Z) > 0$, then $Z$ belongs to the first class $\omega_1$, and if $h(Z) < 0$, then $Z$ belongs to the second class $\omega_2$.

Expected values and dispersions of function $h(Z)$ for each class $\omega_i$ are set by formulae:
\begin{equation} \label{eta-i}
\eta_i =E\{h(Z)\,|\, Z\in \omega_i\} = V^TE\{Z\,|\, Z\in \omega_i\} +v_0 =V^TM_i+v_0, \end{equation}
\begin{equation} \label{sigma-i}
\sigma_i^2 =Var\{h(Z)\,|\, Z\in \omega_i\} = V^TE\{(Z-M_i)(Z-M_i)^T\,|\, Z\in \omega_i\}V =V^T\Sigma_i V.
\end{equation}
For finding optimum values $V$ and $v_0$ a criterion in the form of some function $f(\eta_1,\eta_2,\sigma_1^2,\sigma_2^2)$ is usually used, its critical points define required optimum values. One of widespread criteria is \cite{Fukunaga}, ch. 4, 10:
\begin{equation} \label{widespread-f}
f =\frac{P_1\eta_1^2+P_2\eta_2^2}{P_1\sigma_1^2+P_2\sigma_2^2}.
\end{equation}
This function measures the dispersion between classes (around zero) normalized by dispersion within a class. Then optimum values $V$ and $v_0$  turn out in \cite{Fukunaga}, ch. 4:
\begin{equation} \label{V-v0}
V =[P_1\Sigma_1+P_2\Sigma_2]^{-1}(M_2 -M_1)\quad \text{and} \quad   v_0= -V^T(P_1M_1+P_2M_2).
\end{equation}

\section{Results and discussion}
For creation of the classifying system ECG records of two groups of patients were used. The first group of ECG records of healthy persons contains 96 ECG  fragments for four persons aged from 21 to 27. The second group of ECG records of patients who have recently came through myocardial infarction (subacute period) contains 120 ECG  fragments for five patients aged from 44 to 55. Let us remind that the initial cardiosignals were 30 seconds long with sampling rate of 1028 counts per second. For wavelet decomposition and calculation of features two fragments of signal 8 seconds long each were chosen from each ECG record.

\subsection{Wavelet selecting}
In the wok the orthogonal Meyer wavelet \texttt{dmey} is used, which is derived from Meyer wavelet \cite{Smolen} of infinite impulse response by truncation of its filter to 102 members. It has the carrier on the interval [0,101] and central frequency Fr = 0.6634 Hz. The choice of this wavelet is explained by well localization of frequency spectra of signal components. The point is that this wavelet has the widest frequency spectrum among orthogonal wavelet with the compact carrier. In it the frequencies which are in rather large surrounding area of its center frequency are equally provided 0.6634 Hz. For this reason it provides well expansion of the signal into the items corresponding to certain frequency bands. As ECG sampling rate is 1028 counts per second, in the spectrum of the digitized signal frequencies up to 514 Hz will be provided. Therefore in case of the first level of Meyer wavelets' decomposition it will single out signal elements with the highest frequencies close to center frequency of the first level of decomposition, equal to Fr1 = 0.6634*514 = 340.99 Hz. In case of the second and following levels these frequencies decrease sequentially twice.

\subsection{Wavelet decomposition}
Decomposing of ECG signal to the 4th level is made: $S\mapsto \{D_1, D_2, D_3, D_4, A_4\}$ (Fig. 2). For signal components we have:
$$
S = RecD_1 + RecD_2 + RecD_3 + RecD_4 + RecA_4.
$$
Frequency spectrum of power of the first component $RecD_1$ is concentrated within the limits of 220 to 350 Hz, for the second component $RecD_2$ within the limits of 120 to 200 Hz, for the third component $RecD_3$ -- of 60 to 90 Hz and for the fourth one -- of 25 to 70 Hz (Fig. 3). The choice of the 4th level of decomposition is explained by the fact that low-frequency component $RecA_4$ represents the undistorted smoothed ECG signal cleared of high-frequency oscillations (Fig. 2). In case of more deep expansion of the signal the following high-frequency component $RecD_5$ has frequency spectrum to 30 Hz, and the low-frequency component $RecA_5$ is significantly distorted.
Let us remind that one of the purposes of this work is to show that high-frequency numerical characteristics can be successfully used in ECG classification. Wavelet decomposition of a signal $S$ is made by the following MATLAB command:
\begin{verbatim}
      [c,l] = wavedec(S,4,'dmey');
\end{verbatim}
As a result there is the structure \texttt{[c,l]}, which contains a set of wavelet-coefficients $\{D_1$, \dots, $D_4$, $A_4\}$, where $D_1, D_2, D_3, D_4$ are the coefficients of details and $A_4$ is the approximating coefficients. For reconstruction function \texttt{wrcoef} MATLAB Wavelet Toolbox is used. It allows to restore both high-frequency $RecD_i$ and low-frequency $RecA_i$ components of the signal on the structure \texttt{[c,l]} of wavelet coefficients $\{D_1, D_2, D_3, D_4, A_4\}$:
\begin{verbatim}
for s=1:4
      RecD(s,:)=wrcoef('d',c,l,w,s);
end
\end{verbatim}

\subsection{Feature space and its reduction}
As it was noted in section Feature space, for every high-frequency ECG decomposing component $RecD_1$, $RecD_2$, $RecD_3$, $RecD_4$ it can be calculated to 10 various statistical, frequency and stochastic characteristics. In total there are 40 features for ECG. In the course of work those features which influence was very little (i.e. the corresponding elements of reduction matrix $A$ are small) and those features using of which gave bad results of groups' division have been removed. As a result of such analysis it has turned out that the most suitable feature set for classification aims is the following:
\begin{itemize}
  \item maximum absolute component value;
  \item $L_2$-energy of the component;
  \item maximum value of the power spectrum of the component;
  \item frequency, where maximum value of the power spectrum is achieved;
  \item Shannon's entropy;
  \item Hurst exponent (only for $RecD_4$ component);
\end{itemize}

These features are calculated for every component $RecD_1$, $RecD_2$, $RecD_3$, $RecD_4$. As a result there are 21 features for one ECG record. The specified features form the vector of 21-dimensional space of features. Because of such high dimension it is inconvenient to conduct ECG researches taking into account all features.
The reduction procedure described in section \ref{Reduction} will be applied. Let us calculate scatter matrixes $S_w$ and $S_b$ according to the formulae (\ref{Sw}) -- (\ref{Sb}), then find eigenvalues $\lambda_i$ and eigenvectors $\Psi_i$, $i = 1, \dots, 21$ of ${S_w}^{–1}S_b$  matrix. For this procedure in MATLAB there are \cite{Smolen} function:
\begin{verbatim}
      [Psi,Lambda] = eig(inv(Sw)*Sb).
\end{verbatim}

The eigenvalues \texttt{Lambda} are arranged in descending. Eigenvectors \texttt{Psi} are normalized. According to the procedure stated above it is necessary to choose the subspace of features which is formed by eigenvectors $\Psi_1,\Psi_2,\dots,\Psi_m$ with the greatest values of eigenvalues $\lambda_i$, $i =1,\dots, m$.

As a result of calculations it has turned out that value of the first eigenvalue $\lambda_i$ is approximately equal to $3.4124$, and other eigenvalues have the exponent $10^{-13}$ and lower. That is why the reduced space of features is one-dimensional and formed by eigenvector (column) $\Psi_1$. Reduction matrix $A$ consists of one column, $A = [\Psi_1]$ and affects the full vector of features $Y = [y_1, y_2,\dots, y_{21} ]^T$ as projection $Z = A^TY$ to the one-dimensional space generated by vector $\Psi_1$ of single length. Relative measure of the saved information   at dimension reduction makes approximately 100\%.

During projection of features vectors to one-dimensional space it has turned out that value $Z$ accepts values ranging from  $-2.6\times 10^{-4}$ to $-0.97\times 10^{-4}$. Values' distribution of $Z$ feature is shown on histograms (Fig. 4) separately for the first group of healthy and the second group of sick persons.

\begin{figure}
  \centering
  \includegraphics[width=150mm]{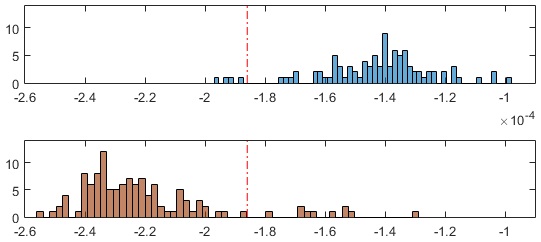}\\
  \caption{The values histograms of $Z$ feature of examined groups of healthy (top) and sick (bottom) persons in the given one-dimensional space of features. The vertical line is determined to be the dividing constant $z_0 = -1.86\cdot 10^{-4}$  of linear classifier.}
  \label{histograms}
\end{figure}


\subsection{Linear classifier construction}
As it was determined in the previous section, the reduced feature space is one-dimensional and can be derived by projection $Z = \Psi_1^T Y$  in one-dimensional space generated by vector $\Psi_1$. The reduced vector of features $Z$ is a scalar. That is why the linear classifier $h(Z) =V^TZ +v_0$ takes the form of $h(Z) = VZ + v_0$, where $V$ is coefficient and $v_0$ is constant term. If $h(Z) >0$, then $Z$ belongs to the first class $\omega_1$, and if $h(Z) <0$, then $Z$ belongs to the second class $\omega_1$. As it comes only to the sign $h(Z)$ it is convenient to study this classifier as
$$
h(Z) = Z + v_0/V = Z -z_0,
$$
where $z_0 = -v_0/V$.  For criterion  by formulae (\ref{estimates}), (\ref{eta-i}) -- (\ref{V-v0}) optimum coefficients $V$ and $v_0$ are calculated and divining constant $z_0$ is found. As a result, $z_0 = -0.0001859$. In Figures 4 and 5 this constant is represented by the vertical red dash-dotted line.

\subsection{Testing of the classifying system}
For testing 96 ECG fragments of four healthy persons aged from 21 to 56 and 120  ECG fragments of five sick persons aged from 45 to 57 who recently came through myocardial infarction (subacute period) are used. For these groups features vectors are created and space reduction of features to one-dimensional space is made, formed by eigenvector $\Psi_1$ for eigenvalue $\lambda_1$. Relative measure of the saved information during reduction of the dimension is approximately 100\%.
For classification the dividing value $z_0= -0.0001859$ received in the previous section is used.

During vectors reduction of features to one-dimensional space it has turned out that value $Z$ take values in the range from от $-2.5\cdot 10^{-4}$ to $-0.8\cdot 10^{-4}$. Values distribution of $Z$ sign for groups of the tested patients is shown on histograms in Figure 5.

\begin{figure}
  \centering
  \includegraphics[width=150mm]{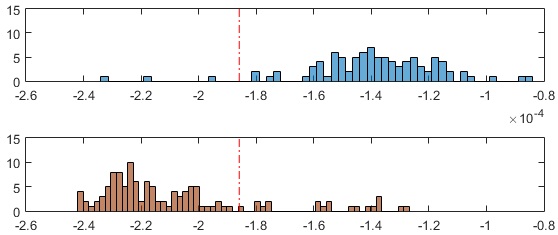}\\
  \caption{The values histograms of $Z$ feature of patients when testing. At the top is features histogram of healthy persons. At the bottom is features histogram of sick persons. The vertical line is the dividing constant $z_0 = -1.86\cdot 10^{-4}$ of linear classifier}
  \label{histograms}
\end{figure}

The data submitted in (Fig. 5) show that only three features (3\%) of 96 given features values of the first group are carried by the classifier to group of sick persons (there are values which are smaller than the dividing constant $z_0$) and only 20 signs (<17\%) of 120 given features values of the second group are carried by the classifier to ECG group of healthy persons (there are values which are bigger than the dividing constant $z_0$).
Let us remind that during ECG record of the patient about 12 assignments are registered. For each assignment the classifier defines whether the ECG assignments belong to ECG-healthy type of persons, or it has some properties typical of the patients who had MI. When testing it has turned out that some healthy persons have assignments features of which are closer to features of sick persons $(h(Z)<0)$. However the average value of all 12 leads of the classifying function $h(Z)$ for all healthy persons participating in testing is higher than zero. The same with sick persons, average value of the classifying function $h(Z)$ of all 12 leads is lower than zero for all sick persons participating in testing. In this sense the classifier is accurate. Besides, it defines "problematic" leads where $h(Z)$ differs from the majority of values in other leads. For every patient the set of 12 classifier values $h(Z)$ in every lead can serve as additional characteristics of patient's condition.

For example, values of qualifier $h(Z) = Z -z_0$ are calculated for 12 standard leads of the personal cardiogram of the first author of this work (the values multiplied by 103 are given below):
\begin{verbatim}
–0.0561  –0.2441   –0.7162    0.5211   –0.0255   –0.5415    0.1347   –0.1841
–0.0558   –0.4035    0.5141  –0.3891.
\end{verbatim}
The classifier defines ECG problems in 9 leads. Therefore this patient (the first author) does not belong to the group of healthy persons, and it is true. At the same time, the average value equal to $-0.1205\times 10^{-3}$ is close to zero, therefore there are no bases to put this patient into the second group of the persons who had MI. Also, the cardiogram of the second author is studied. In this case values of the classifying function for all 12 leads of ECG are positive. The second author is attributed by the qualifier to "healthy" group, and it is also true.

\section{Conclusion}
Based on the wavelet analysis, a classification system has been constructed that reliably separates groups of healthy and sick patients.
Positive results of testing show that high-frequency ECG wavelet-components carry essential diagnostic information concerning ECG.
Such a system can be used as a complement to classification systems based on analysis of the wave complexes and PQRSTU segments for more accurate and informative separation of patient groups.


\begin{thebibliography}{999}
\bibitem{Adeli}
{\it Adeli H., Ghosh-Dastidar S., Dadmehr N.} A Wavelet-Chaos Methodology for Analysis of EEGs and EEG Sub-Bands to Detect Seizure and Epilepsy. IEEE Transactions on Biomedical Engineering, 2007, vol. 54(2), pp. 205--211.

\bibitem{Aldroubi}
{\it Aldroubi A., Unser M.} Wavelets in Medicine and Biology. CRC Press, 1996.

\bibitem{Daube}
{\it Daubechies, I.} Ten Lectures on Wavelets, SIAM, Philadelphia, 1992.

\bibitem{Fukunaga}
{\it Fukunaga  K.}  Introduction  to  Statistical  Pattern  Recognition. Academic  Press, Boston, 1990.

\bibitem{Ishikawa-02}
{\it Ishikawa Y., Mochimaru F.} Wavelet Theory–Based Analysis of High-Frequency, High-Resolution Electrocardiograms: A New Concept for Clinical Uses. Progress in Biomedical Research, 2002, vol. 7, no. 3, pp. 179--184.

\bibitem{Ishikawa}
{\it Ishikawa Y.} Wavelet Analysis for Clinicial Medicine. Chapter 6: SAECG (Signal Averaged ECG) which was seen from Wavelet Analysis – Supplement – original color images. Available at: http://www.uinet.or.jp/~ishiyasu/ch6/index.html.

\bibitem{Mochimaru}
{\it Mochimaru F., Fujimoto Y.}  Detecting the Fetal Electrocardiogram by Wavelet Theory-Based Methods. Progress in Biomedical Research, 2002, vol. 7, no. 3, pp. 185--193

\bibitem{Nagendra}
{\it Nagendra H., Mukherjee S., Kumar V.} Application of Wavelet Techniques in ECG Signal Processing: An Overview. International Journal of Engineering Science and Technology (IJEST), 2011, vol. 3, no.10,  pp. 7432--7443.

\bibitem{Pavlov}
{\it Pavlov A.N., Khramov A.E., Koronovskiy A.A., Sitnikova E.Yu., Makarov V.A., Ovchinnikov A.A.} Wavelet analysis in neurodynamics. UFN, 2012, vol. 182, no. 9, pp. 905--939. (In Russian).

\bibitem{Sambhu}
{\it Sambhu D., Umesh A. C.} Automatic Classification of ECG Signals with Features Extracted Using Wavelet Transform and Support Vector Machines. IJAREEIE, 2013, vol. 2, no. 1, pp. 235--241.

\bibitem{Shahbahrami}
{\it Shahbahrami A., Kiani M.} Classification of ECG arrhythmias using discrete wavelet transform and neural  networks. IJCSEA, 2012, vol.2, no.1, pp. 1--13.

\bibitem{Smolen}
{\it Smolentsev N.K.} Fundamentals of the theory of wavelets. Wavelets in MATLAB. Moscow: DMK Press, 2013. (In Russian).

\bibitem{Smolen-ML}
{\it Smolentsev N.K.} MATLAB. Programming in C\#, Java and VBA. Moscow: DMK Press, 2015. (In Russian).

\bibitem{Vorobyov}
{\it Vorobyov A.~S.} Electrocardiography. The newest guide. Moscow: Publ. house Eksmo; Saint Petersburg: Sova Publ., 2003. (In Russian).

\end{thebibliography}
\end{document}